\documentclass[12pt,a4paper]{article}
\usepackage[margin=1in]{geometry}

\usepackage[affil-it]{authblk}
\usepackage[version=3]{mhchem} 
\usepackage[T1]{fontenc}       
\usepackage{array}

\usepackage{mathrsfs,amsmath,amssymb}
\usepackage{graphicx,color}
\usepackage{subcaption}
\usepackage{makecell}
\usepackage{comment}
\usepackage{etoolbox}
\usepackage{float}
\usepackage[dvipsnames]{xcolor}
\usepackage[normalem]{ulem}
\newcommand*{\dd}{\mathrm d}
\newcommand*{\ww}{\mathrm w}
\newcommand*{\hh}{\mathrm h}
\newcommand*{\aaa}{\mathrm a}

\newcommand*{\sGL}{\sigma_{\scriptscriptstyle\rm GL}}
\newcommand*{\sSL}{\sigma_{\scriptscriptstyle\rm SL}}
\newcommand*{\sSG}{\sigma_{\scriptscriptstyle\rm SG}}
\newcommand*{\sij}{\sigma_{\scriptscriptstyle\rm ij}}
\newcommand*{\Jij}{J_{\scriptscriptstyle\rm ij}}

\newcommand*{\ttc}{\theta_{\scriptscriptstyle\rm C}}
\newcommand*{\tty}{\theta_{\scriptscriptstyle\rm Y}}
\newcommand*{\tth}{\theta_{\rm H}}

\newcommand*{\aL}{a_{\scriptscriptstyle\rm L}}
\newcommand*{\nN}{n_{\scriptscriptstyle\rm neigh}}
\newcommand*{\DF}{\Delta \mathcal{F}}
\newcommand*{\SCB}{S_{\scriptscriptstyle\rm CB}}

\newcommand*{\BCB}{B_{\scriptscriptstyle\rm CB}}
\newcommand*{\SW}{S_{\scriptscriptstyle\rm W}}

\newcommand*{\BW}{B_{\scriptscriptstyle\rm W}}
\newcommand*{\phiS}{\phi_{\scriptscriptstyle\rm S}}
\newcommand*{\ALGcap}{A_{\scriptscriptstyle\rm GL}^{\scriptscriptstyle\rm C}}
\newcommand*{\ALGbot}{A_{\scriptscriptstyle\rm GL}^{\scriptscriptstyle\rm B}}
\newcommand*{\ALG}{A_{\scriptscriptstyle\rm GL}}
\newcommand*{\ALS}{A_{\scriptscriptstyle\rm LS}}
\newcommand*{\ASG}{A_{\scriptscriptstyle\rm SG}}
\newcommand*{\AVS}{A_{\scriptscriptstyle\rm VS}}

\begin{document}

\title{Can one predict a drop contact angle?}

\author{Marion Silvestrini$^{*,1,2}$, Antonio Tinti$^2$, Alberto Giacomello$^2$, Carolina Brito$^1$  \\
	$^{1}$Instituto de Física, Universidade Federal do Rio Grande do Sul, Av. Bento Gonçalves 9500, CEP 91501-970, Porto Alegre, Brazil   \\
	$^{2}$Dipartimento di Ingegneria Meccanica e Aerospaziale, Università di Roma ‘‘La Sapienza’’, 00184 Rome, Italy \\
	}

\maketitle


\begin{abstract}

The study of wetting phenomena is of great interest due to the multifaceted technological applications of hydrophobic and hydrophilic surfaces. The theoretical approaches proposed by Wenzel and later by Cassie and Baxter to describe the behaviour of a droplet of water on a rough solid were extensively used and improved to characterize the apparent contact angle of a droplet. However, the equilibrium hypothesis implied in these models means that they are not always predictive of experimental contact angles due to strong metastabilities typically occurring on heterogeneous surfaces. 
A predictive scheme for contact angle is thus urgently needed both to characterise a surface by contact angle measurements and to design superhydrophobic and oleophobic surfaces with the desired properties, e.g., contact angle hysteresis. In this work a combination of Monte Carlo simulation and the string method is employed to calculate the free energy profile of a liquid droplet deposited on a pillared surface. 
For the analyzed surfaces, we show that there is only one minimum of the free energy that corresponds to the superhydrophobic wetting state while the wet state can present multiple minima. Furthermore, when the surface roughness decreases the amount of local minima observed in the free energy profile increases.

\end{abstract}

$^{*}$\texttt{email: marion.silvestrini@ufrgs.br}

\section{Introduction}
Contact angle is perhaps the most immediate way to characterise the wettability of a given surface, due to the ease of depositing a drop and estimate the angle formed by it. Indeed, the Young equation allows one to relate the surface tensions with the contact angle on an ideally smooth surface \cite{Young1805}. However, the crucial question is whether this observable is a reliable one in the case of actual surfaces, in which heterogeneities are inevitably present.

Surface heterogeneities can dramatically change the contact angle as compared to the case of an ideal, homogeneous surface. In particular, topographical heterogeneities tend to magnify the intrinsic wettability of the surface, making them ``superhydrophobic'' or ``superhydrophilic'' \cite{Quere2008}. Since the pioneering work of Wenzel \cite{Wenzel1936} and Cassie and Baxter \cite{Cassie1944}, the concept of \emph{apparent} contact angle has been used, which is meant to predict the angle measured at the actual surface accounting for the effect of both wettability and heterogeneity. Since then, a continued effort has been spent to provide accurate models for predicting the apparent contact angles based on the chemical and topographical characteristics of the surface. 
The main ambition of this line of research is finding a rapid and inexpensive way to characterise surface wettabilities via contact angle measurements.

The standard approach to develop wetting models for heterogeneous surfaces consists in assuming a wetting state of the surface and predict the contact angle based on the homogenised surface energy of such composite interface. The task becomes more involved when a surface allows multiple visibly different states, e.g., for superhydrophobic surfaces in which the roughness can be fully wet (Wenzel, W) or dry (Cassie-Baxter, CB).

Importantly, major difficulties arise in relating the apparent contact angle and surface characteristics, most importantly due to \emph{static} contact angle hysteresis (CAH), i.e., a scatter of the contact angle values which depends on the experimental procedure or history. Already in 1964, Johnson and Dettre \cite{johnson1964} showed that an idealised sinusoidal roughness can be an important source of CAH; they further showed that the W state is connected to larger CAH than the CB state. More recent work further showed that CAH emerges from micro and nanoscale phenomena which are beyond the scope of Cassie and Wenzel equations \cite{Nosonovsky2007}.
A heated debate originated from the question whether ``Cassie and Wenzel were wrong'' \cite{gao2007,Marmur_CWright,mchale2007}. The main merit of such discussion was to highlight that the experimental contact angle may be determined by features localised at the three-phase contact line where the drop meets the surface. In fact, the classical CB and W models assume an equilibrium hypothesis in which the droplet is allowed to relax to the absolute free energy minimum. In other words, due to the rough free energy landscape featuring multiple minima, the measured contact angle may significantly depart from the value predicted based on averaged surface characteristics and the commonly used wetting models \cite{erbil2009,erbil2014,erbil2021}. 

Several attempts have been made to relate the characteristics of single defects to CAH, starting from the pioneering work of Joanny and de Gennes \cite{joanny1984,reyssat2009,semprebon2012,giacomello2016wetting}. However, densely packed heterogeneities \cite{crassous1994}, together with the finite size of the drop and the multiple wetting states possible on superhydrophobic surfaces \cite{yan2020wetting,li2005,Kusumaatmaja2007,schellenberger2016,patankar2010,ren2014} considerably complicates the matter. Indeed to date, notwithstanding the theoretical and practical importance, the analysis of the free energy landscape connected to wetting of a realistic surface by a droplet is still lacking.
Thus making a connection between measured contact angles and available wetting models is highly needed in the community. At the same time, the difficulties connected to the exploration of rough free energy landscapes \cite{bonella2012} represent a major obstacle to develop the computational tools required for engineering textured surfaces with the desired wetting properties. In this work, we fill such a gap, providing a generic framework for reconstructing the free energy connected to wetting of a pillared superhydrophobic surface based on the combination of the cellular Potts model \cite{Graner1992} and the (zero temperature) string method for rare events \cite{e2002,maragliano2006,e2007}. This approach allows us to assess the variation of CAH with the pillar spacing, to connect CAH  with the existence of multiple free energy minima, and to verify the energetic origin of the superhydrophobicity of the CB state. In addition, the presented approach provides useful guidelines for contact angle measurements on actual surfaces and for designing surfaces with tailored wetting properties, notably superhydrophobic ones.

The paper is organised as follows. In \textbf{Section~\ref{sec:numerics}} we present the numerical setup used: a Monte Carlo scheme combined with the string method. \textbf{Section~\ref{sec:results}} contains the results of the free energy calculations and the physical origin of its saddle points. In \textbf{Section~\ref{sec:discussion}}, we discuss the origin of contact angle hysteresis and compare the CB and W models with the simulations. The paper is then concluded in \textbf{Section~\ref{sec:conclusions}}.


\section{Numerical Model}
\label{sec:numerics}
\begin{figure}[H]
\centering
\includegraphics[width=.6\linewidth]{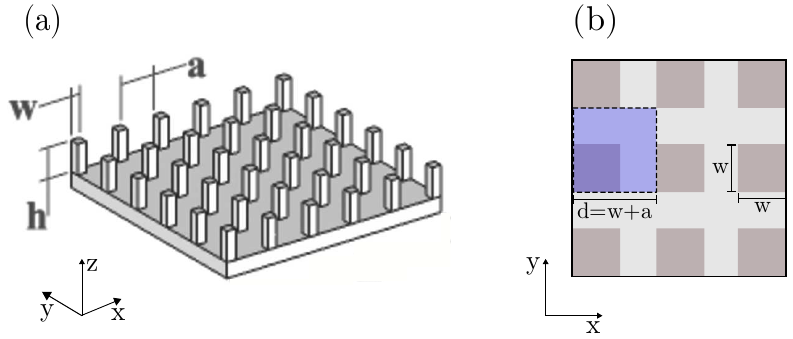}
\caption{Sketch of the surface decorated with pillars. \textbf{(a)} Geometrical parameters defining the texture: $\ww$ is the pillar width, $\hh$ is the pillar height, $\aaa$ is the interpillar distance and $\dd = \ww + \aaa$. \textbf{(b)} Top view of the surface. The blue region delimits one cavity, used in the definition of the collective variables.} 
\label{typesSurface}
\end{figure}

The main aim of our simulation campaign is connecting the apparent contact angles attainable on a given surface with its physical and chemical characteristics. We focus on a hydrophobic surface decorated with a three-dimensional  array of pillars with the geometry shown in Figure~\ref{typesSurface}-a. 
A three-dimensional spherical droplet with volume  $V_0=4/3\pi R_0^3$ is placed over the surface. We use a mesoscale wetting model, the cellular Potts model \cite{Graner1992,Dan2005,Oliveira2011}, which has been successfully used in studying the wetting of structured surfaces \cite{Lopes2013, Mortazavi2013,Fernandes2015,xu2021,lopes2017}. Its low computational cost as compared to, e.g., molecular dynamics, allows computations of relatively large 3D drops via Monte Carlo minimisation. 

We combine the MC minimisation with the string method \cite{e2002,e2007} in order to calculate the most probable path leading from the superhydrophobic CB to the fully wet W state or vice versa. The string method, combined with atomistic simulations, diffuse interface methods, or sharp interface models, has proven instrumental to clarify the origin of metastability in wetting of structured surfaces \cite{ren2014,zhang2014,pashos2015,giacomello2015}. We will employ this approach to overcome the free energy barriers, to establish the mechanism of breakdown/recovery of the CB state, and to identify intermediate minima. 

\subsection{The cellular Potts model}

The Cellular Potts model (CPM) assumes a three state system in a simple cubic lattice, with each site representing a different state: solid, liquid, and gas. We employ this technique to simulate a three-dimensional droplet in contact with air and a solid surface. The Hamiltonian of the system is given by:

\begin{equation}
H = \frac{1}{2} \sum \limits_{\langle {\rm i,j} \rangle} E_{s_{\rm i},s_{\rm j}} \left( 1 - \delta_{s_{\rm i},s_{\rm j}} \right) + \frac{\kappa}{2} \sum \limits_{l=1}^{l=N} \left( z_l - z_l^T \right)^2,
\label{hamilton}
\end{equation}
where $s_{\rm i} \in \{0, 1, 2\} $ is the state of a given site in the lattice, gas (G), liquid (L), or solid (S). The first term on the right hand side describes the surface interaction between the three states, with ${E_{s_{\rm i},s_{\rm j}} = \Jij \cdot \aL}$, where $\Jij$ is the site-site coupling  between $i$ and $j$ type sites, $\aL=L^2$ is the unit area and $L$ is the lattice size. The summation ranges over the $\nN =26$ nearest and next-nearest lattice sites of the 3D simple cubic lattice. Notice that the energy contribution will be non-zero only when $s_{\rm i} \neq s_{\rm j}$. Gravitational contributions to the energy are neglected due to small size of the droplets hereby investigated~\cite{Shahraz2012,xu2021,Fernandes2015}. 

The second term in Equation~\eqref{hamilton} corresponds to a harmonic bias used to drive the system across metastabilities allowing to more effectively explore phase space \cite{maragliano2006} using general descriptors, the \emph{collective variables}  $z$, which are function of the lattice sites. Biasing is used because, in an {\it unrestrained} simulation ($\kappa=0$), drops relax and remain trapped in the closest minimum state. Simulation modeling a droplet deposited on a textured surface without any bias were previously studied in Refs.~\cite{Fernandes2015,silvestrini2017}, where the authors indeed showed that the final contact angle depends on the initial wetting configuration.

The collective variables $z$ that appear in Equation~\eqref{hamilton} should be chosen carefully \cite{amabili2018} in order to correctly discriminate among relevant configurations of the system. More in detail $z^T$ represents the target value for the collective variable, while $\kappa$ is stiffness of the harmonic constraint. In this work we shall identify $z_l$ with liquid-phase occupation of $N$ interpillar grooves. The usage of such discrete density indicators is customary when dealing with wettability problems \cite{tinti2017}.

The energy scale is set once $\Jij$ is defined. In this work spin-spin couplings are chosen to render the same ratios between the interfacial surface tensions as in Ref.~\cite{Tsai2010}, in which a droplet of water is studied on a surface composed by Polydimethylsiloxane (PMDS) with the following surface tensions $\sij$ between the gas (G), liquid (L), and solid (S): $\sGL = 70 \times 10^{-3}$ N/m, $\sSG = 25 \times 10^{-3}$ N/m, and  $\sSL=53 \times 10^{-3}$ N/m; the last value was calculated by Young's equation, $\sGL \cos\tty = \sSG - \sSL$ with $\tty=114^{\circ}$. 
The data will be presented in dimensionless form using the lattice size $L$ as unit length, $\aL = L^2$ as unit area, and $J_\mathrm{GL} \cdot \aL$ as unit energy, referred as {\it e.u}. 
In such reduced units, $J_\mathrm{GL}=1$, $J_\mathrm{SG}=0.36$, and $J_\mathrm{SL}=0.76$.  
For the forcing parameter we use $\kappa\simeq 0.004$,
which drives the system to the target configuration while allowing some fluctuation. More details about the choice of $\kappa$ is reported in the \emph{Supporting Information} (SI).
In order to convert the dimensionless units to physical ones, one can for instance assume $L=1 \, \mu$m and the energy for the gas-liquid interface of water, which yields $\sGL \aL=7 \times 10^{-14}$ J. 

To evolve the system we use the Metropolis-Hastings algorithm, which consists in changing the state of two random sites at the gas-liquid interface with an acceptance rate equal to min$ \{1, {\rm exp} [- \beta \Delta H ] \} $, where $\beta=1/T$ is the inverse of the effective temperature of the CPM, which introduces some noise, allowing for a more effective exploration of the phase space. We set $T= 4.8$ throughout the article, which allows the  system to fluctuate with an acceptance rate of approximately 9$\%$.
The attempted MC moves consist in swaps between liquid and gas sites, which guarantees that the volume of the droplet is constant throughout the simulation. 

\subsection{The string method}

In order to attempt a simulation of the wetting of rough surfaces it is crucial to tackle the challenge associated with the presence of metastabilities.
Previous methods to sample rough free energy landscapes required projecting to a low dimensional representation of the free energy as a function of a handful of parameters and reconstructing the full landscape as a function of such collective parameters. It is easy to understand how this task requires a computational effort which is exponential in the number of collective variables thus limiting the applicability of this class of methods typically to two/three variables. Such low dimensional representation of the energetic landscape often results in a poor description of the phenomena \cite{amabili2018}.
The (zero temperature) string method in collective variables was first introduced by Vanden-Eijnden and collaborators \cite{maragliano2006} and is a path method that requires only the computation of the \emph{local} gradient of the free energy landscape at certain points along the path; its computational cost thus only grows linearly with the number of collective variables.

The string method allows for a fast and convenient identification of the minimum free energy path (MEP) connecting two metastable regions of the rough landscape, along with providing a measure of the free energy along the path, without requiring to fully sample a high dimensional variable space. This is achieved by iteratively refining a discretized guess path (i.e., the string).
In this work the collective variables (CV) represent the liquid occupation of each cavity, as represented by the blue region in Figure~\ref{typesSurface}-b. The available volume in the cavity is given by $(\dd^2- \ww^2) \hh$.
The algorithm proceeds as follows:
\begin{enumerate}
    \item We run $N_r$ Monte Carlo (MC) replicas, each biased to explore the vicinity of a $z^T_{l,i}$ point in the collective variable space.  Each replica corresponds a string point in the collective variable space.
    \item The $N_r$ replicas run for $10^5$ MC moves allowing to sample the mean biasing forces and the local free energy gradient at each string point.
    \item The string is updated by a convenient gradient descent of the string points and reparametrized \cite{e2007} so that one has a new set of $z^T_{l,i}$. The process is iterated from point {\bf 1}.
\end{enumerate}

Refinement of the initial string guess is repeated until convergence is reached, as evaluated by a suitable threshold in the changes to the path/path energy. In all runs convergence was obtained within 20 iterations. At convergence, numerical integration of the free energy gradients allows for the reconstruction of the free energy profile along the string:
\begin{equation}
\Delta \mathcal F_j=  \sum_{i=1}^j \sum_{l=1}^{l=N} - \kappa \langle (z_{l,i}- z^T_{l,i})\rangle \Delta z^T_{l,i} \text{ ,}
\label{eqFE}
\end{equation}
with $\Delta \mathcal F_j$ the free energy at point $z^T_{l,j}$, the index $l$ running over the $N$ collective variables, $i$ up to the current replica $j$, with $j\leq N_r-1$, and 
$\Delta z^T_{l,i}=z^T_{l,i+1}-z^T_{l,i}$. 
The results shown in this work are averages over the distinct realizations of the simulation and $\langle \dots \rangle$ represents the average over MC steps.


\section{Results}
\label{sec:results}

We consider three substrates with same pillar height $\hh=10L$ and width $\ww=5L$ and different distances between pillars: substrate referred to as S$_1$ presents a low interpillar distance ($\aaa=5L$), S$_2$ an intermediate value ($\aaa=8L$), while S$_3$ has the largest interpillar distance ($\aaa=11L$). The volume of the spherical droplet is defined by imposing an initial radius of $R_0=50L$.
We first show the rough free energy landscape connected with wetting of the three surfaces, characterizing the droplet configuration at the minima and maxima of the free energy. We then introduce a possible physical explanation of the roughness in the free energy and end this section discussing the connection between the free energy landscape and contact angle hysteresis.

\subsection{Rough free energy of a hydrophobic pillared surface}

\begin{figure}[H]
\centering
\includegraphics[width=.6\linewidth]{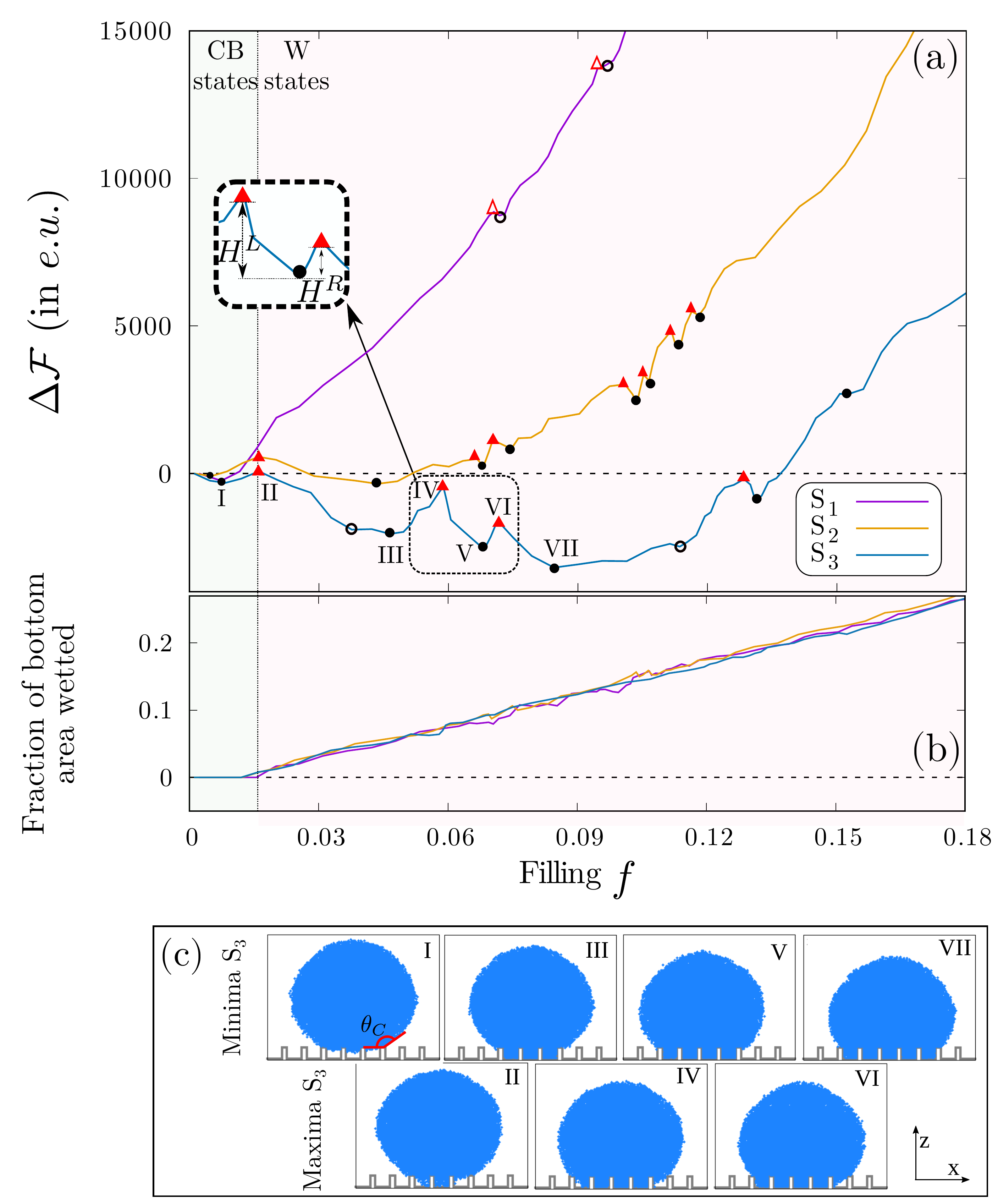}
\caption{\textbf{(a)} Free energy profile as a function of the liquid filling the cavities for the three substrates. Minima are identified with black circles and maxima by red triangles; empty symbols denote very shallow minima (and related maxima) in which the system does not remain after standard MC minimisation. \textbf{(b)} Fraction of the bottom area of the substrate in contact with the liquid, normalized by the total substrate area. \textbf{(c)} Lateral view of the minima and maxima in the wetting of substrate S$_3$. The inset in \textbf{(a)} defines the  left and right barriers, $H^L$ and $H^R$, respectively.}
\label{FE}
\end{figure}

Figure~\ref{FE}-a reports, for substrates S$_1$, S$_2$, and S$_3$, the free energy of the drop as a function of the total filling of the cavities below the drop, which is defined as the sum of the target filling of each cavity divided by the total volume of the droplet: $f =\sum_l z_l^T/V_0$. 
The minima in the free energy correspond to stable (global) and metastable (local) states of the system. The free energy extrema are numerically identified from the data; minima are indicated in Figure~\ref{FE}-a as black circles, while maxima by red triangles.
Since some minima are very shallow and may be affected by numerical accuracy 
we performed numerical MC minimisation using different initial conditions close to the putative minima, to confirm that the system can indeed reside in such states; empty symbols denote shallower minima which will not be further discussed.
Figure~\ref{FE}-b shows the fraction of water wetting the bottom area.
The vertical dotted line in the figure 
roughly separates two regimes: on the left, the liquid does not touch the bottom of the substrate, indicating that the corresponding configurations of the droplet are associated to the superhydrophobic CB state, while, on the right of the line, the liquid reaches to the bottom of the substrate and the droplet configurations are associated with the wet W state(s). Importantly, the free energy landscape appears rough for all considered surfaces, with markedly different trends for the three interpillar distances: monotonically growing for S$_1$, almost at the CB-W coexistence, with additional high free energy minima,  for S$_2$, and with a significant W basin with multiple local minima for S$_3$.

Each point on the free energy profiles corresponds to a droplet configuration, whose sequence thus defines a wetting path, see Figure~\ref{FE}-c; in particular, these paths represent the most probable way in which the transition from CB to one of the various W states (or vice versa) occurs. Other paths may exist connecting minima, especially in such complex landscape, see e.g. Ref.~\cite{tinti2018}. 
Figure~\ref{FE}-c presents a lateral view of the 3D droplet in correspondence of the minima and maxima of substrate S$_3$. Minimum I corresponds to the CB state, with air trapped between the pillars -- this numbering is the same for the three substrates. As the droplet starts infiltrating the substrate, it touches the bottom for the first time at point II, where free energy increases to a maximum. A similar behaviour was also reported in molecular dynamics simulations \cite{ren2014,Escobedo2012a,Escobedo2012b,amabili2017}.
As the filling level increases, the free energy of the substrate presents several local minima which correspond to the progressive infiltration of liquid within the pillars: lateral views of the droplet in Figure~\ref{FE}-c clearly show full wetting of 3 lines of cavities at minimum III, 4 lines  at V, 5 lines at VII. 
For filling levels above 7 lines of cavities, the free energy does not present any other minima for this droplet size. 

The maxima separating the mentioned minima are found to be associated to an incomplete wetting of some of the cavities at the drop perimeter; we will further analyse their origin in Section~\ref{physOrigin}. We note that the wetting of the substrate in 3D is a more complex problem than what can be inferred from a lateral view, which is however convenient to picture the main features of the process; we thus make available  in the \emph{Supporting Information} videos in 2D and 3D of the droplet wetting the 3 substrates.

The inset of Figure~\ref{FE}-a defines the barrier $H^L$, which is the difference in $\DF$ between a minimum and the consecutive maximum on its left and $H^R$, being the difference in $\DF$ between a minimum and the first maximum on its right. Using these definitions, we found that the barrier $H^R$ and $H^L$ are typically of the same order of magnitude for substrate S$_3$, while for S$_1$ $H^L \ll H^R$. We will come back to this point later. 

These observations raise several questions, which are addressed in the next section: why some substrates present global minima at the CB state and others in the W configuration? What is the physical origin and significance of the local minima and the intervening maxima?

\subsection{Physical origin of the minima and maxima of the free energy}
\label{physOrigin}

The wetting state corresponding to the global free energy minimum is different for the considered substrates and can be rationalized using a continuous global energy model \cite{Fernandes2015}. This model takes into account the energy of creating interfaces between the liquid, the gas, and the solid  of a  spherical droplet with fixed volume $V_0$ placed on a surface. The droplet is allowed to display two wetting states: it either stays in the superhydrophobic CB state or it {\it homogeneously} wets the surface in the W state. We refer to these configurations as ideal Cassie-Baxter, CB$^I$ and ideal Wenzel, W$^I$, states. 
The difference in energy of a system with and without the droplet on the surface can be written as \cite{Fernandes2015}: 
\begin{eqnarray}
\Delta E^{\bf {CB^I}} &=&  \sGL ~\left[~ \SCB - \pi \BCB^2 \left( \phiS \cos\tty - (1-\phiS) \right) ~ \right],\label{en_CB}\\
\Delta E^{\bf {W^I}} &=& \sGL ~\left[~ \SW - \pi \BW^2 \, r \cos\tty~ \right] ,\label{en_W}
\end{eqnarray}
where $S_{\rm s}=2\pi  {R^{\rm s}}^2 \left[ 1-\cos (\ttc^{\rm s} ) \right] $ is the surface of the spherical cap in contact with air, $B^{\rm s} = R^{\rm s}\sin(\ttc^{\rm s})$ is the base radius, $R^{\rm s}$ the radius of the droplet, and $\ttc^{\rm s}$ its contact angle in the state $s$. $ \phiS= \ww^2/\dd^2$ is the fraction of solid surface area wet by the liquid (or pillar density) and $r=1+4\ww\hh/\dd^2$ is the surface roughness ratio.

The main difference between the model defined by Equation~\eqref{en_W} and the Wenzel one~\cite{Wenzel1936} is that the former one considers a droplet of finite size, which means that the interface of the cap and the lower interface in contact with the substrate compete in the minimisation of the total free energy. We will return to this point in Section~\ref{sec:discussion}. Concerning the CB state, it is instead found that results are equivalent when one explicitly considers the cap as in Equation~\eqref{en_CB} or simply homogenises the surface free energy as in the original CB theory \cite{Cassie1944}.

\begin{figure}[H]
\centering
\includegraphics[width=.6\linewidth]{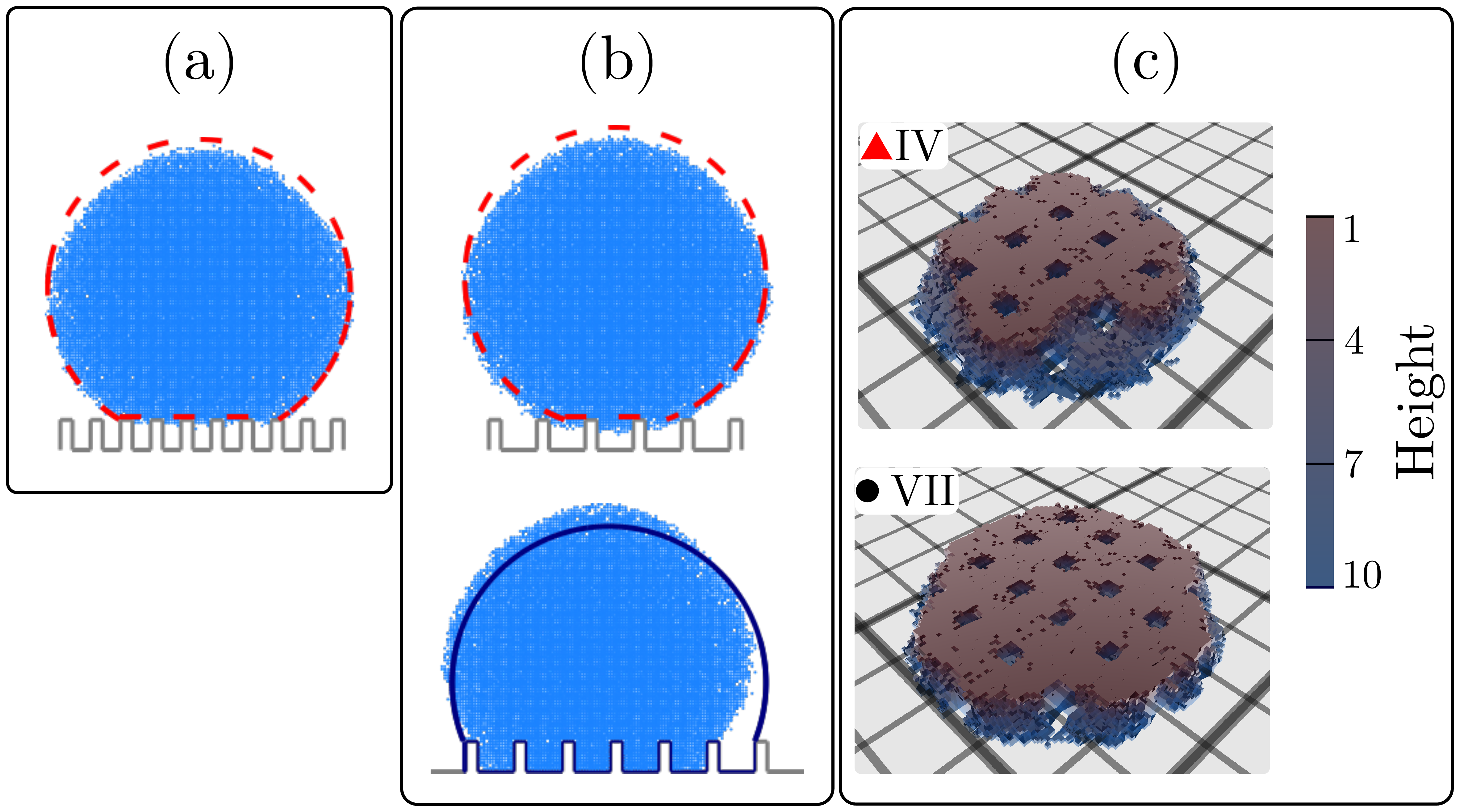}
\caption{Homogenised model {\it vs} $\DF$. {\bf(a)} Side view of  configuration at the minimum of $\DF$ for surface S$_1$ in blue. Red dashed line is the numerical solution of the homogenised model for the CB$^I$ state, Equation~\eqref{en_CB}. {\bf(b)} Top: side view of a sessile droplet on surface S$_3$ at the local minimum of $\DF$ (point I in Figure~\ref{FE}) and the solution of Equation~\eqref{en_CB}. Bottom: global minimum of $\DF$ in the W state (point VII in Figure~\ref{FE}). The blue solid line is the numerical solution of the model in Equation~\eqref{en_W} for the W$^I$ state.
{\bf(c)} View from below of the 3D configuration of the droplet placed on S$_3$ at two filling levels: point IV in Figure~\ref{FE}-a (maximum) and point VII (global minimum); colors correspond to the height of liquid in each lattice site.}
\label{ContFE}
\end{figure}

Figures~\ref{ContFE}-a and b show lateral views of droplets minimising the equations of the homogenised model and $\DF$ obtained by MC simulations for surface S$_1$ and S$_3$, respectively. 
$\DF$ of the surface S$_1$ presents only one minimum for which the simulated configuration is shown in blue together with the minimum of the solution of Equation~\eqref{en_CB}, in red. 
$\DF$ of surface S$_3$ has multiple minima and two of them are represented in Figure~\ref{ContFE}-b: the local minimum at the CB state, shown on top with the solution of Equation~\eqref{en_CB}, and the global W minimum reported in the bottom panel, in which is compared with the solution of Equation~\eqref{en_W}.
Results are in good agreement for the CB state, both concerning the contact angle and the number of intruded cavities. On the other hand, contact angles are not in perfect agreement for the W state, due to the local pinning at pillars, which is not captured in Equation~\eqref{en_W}, see below for details. On the other hand, the model predicts correctly that W is the global free energy minimum and the free energy values are in reasonable accord, see the \emph{Supporting Information}.

Besides the fair agreement at the global minimum, the main difference between the two approaches is that the simulated free energy profile presents several local minima which are not accounted for in any homogenised model. This difference is particularly important for the W state, which helps to explain the difference in the global W minimum. 
Indeed, an important simplification introduced in the model~\eqref{en_W} is that the W state is achieved by \emph{homogeneous} wetting: below the droplet base, there is a perfect cylinder filling the cavities; individual wetting of pillars is thus disregarded; the cost to create the interface between the liquid and air is also neglected in the model. However, simulations show that the infiltration of the substrate is not homogeneous, that pinning at individual pillars may occur, and that the interface between the liquid and the gas below the droplet is quite rough, as exemplified in Figure~\ref{ContFE}-c for one minimum and one maximum: the configurations are particularly far from being a cylinder when $\DF$ is at a maximum.

The nontrivial shape of the droplet in contact with the surface drove us to investigate the contribution to the free energy of each interface. We propose a putative free energy defined as: $\Omega = \Delta P \,V_0 + \sGL \ALG +\sSL \ALS + \sSG \ASG$, where $\Delta P$ can be interpreted as the Laplace Pressure,  $\Delta P = -\alpha \sGL$ with $\alpha$ being the mean curvature of the drop and $A_{ij}$ are the interfaces between two different phases ${ij}$ (solid, liquid, and gas).
We then use the Young equation $\sSG = \sSL + \sGL \cos \tty$ to rearrange the terms and note that the total area of the substrate $A_\mathrm{tot}=\ALS + \AVS$ is constant.  
The contribution of liquid-gas surface area $\ALG$ is split in two parts, one corresponding to the surface of the spherical cap $\ALGcap$, and other to the part below the droplet, $\ALGbot$. We can then write the difference in free energy with respect to the reference one $\Omega_\mathrm{ref}=\sSG A_\mathrm{tot}$: 
\begin{equation}
\Delta \Omega = \sGL \left(\ALGcap - \alpha V_0 ~+~ \ALGbot - \ALS\cos\tty\right).
\label{DOmega}
\end{equation}
Interestingly, the surface tension $\sGL$ factorises and only geometrical quantities appear in the parenthesis on the right hand side, together with the contact angle $\tty$.

\begin{figure}[H]
\centering
\includegraphics[width=.7\linewidth]{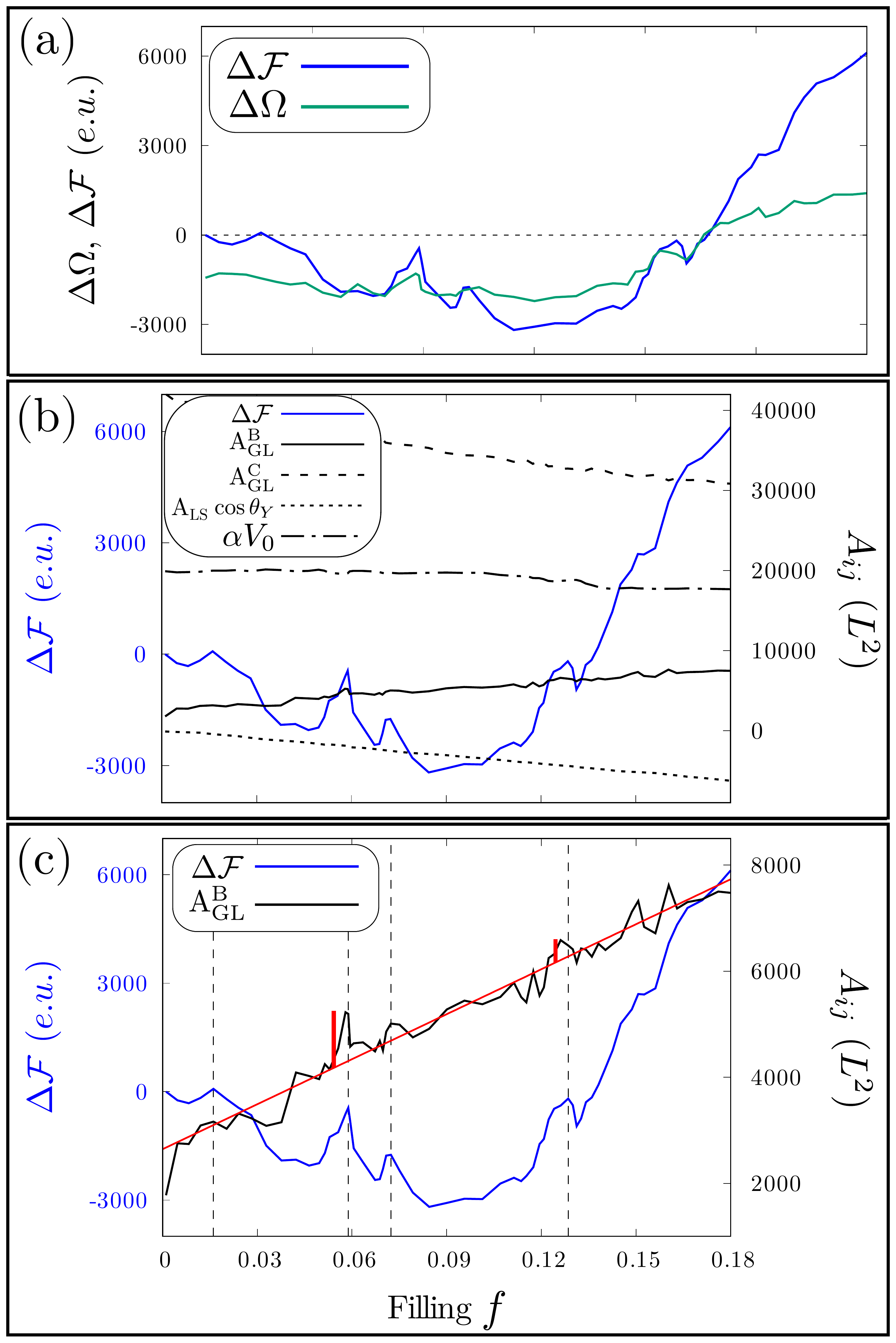}
\caption{$\DF$, $\Delta \Omega$ and areas as a function of $f$ for S$_3$. {\bf (a)} Comparison of $\DF$ and $\Delta \Omega$. {\bf (b)} Each geometrical term in Equation~\eqref{DOmega} is shown separately and compared with $\DF$ in blue. {\bf (c)} $\DF$ in blue and $\ALGbot$ is shown in black. The red curve is a linear fit of $\ALGbot$ and the red vertical lines indicate the area variations discussed in the main text. Dashed lines denote the maxima of $\DF$.}
\label{areasFE}
\end{figure}

Figure~\ref{areasFE}-a compares, for surface S$_3$, the free energy $\DF$ obtained in the MC simulations via Equation~\eqref{eqFE} with the putative free energy $\Delta \Omega$ always computed from MC simulations but by measuring the geometrical quantities in Equation~\eqref{DOmega}. 
In particular, the surface areas $A_{ij}$ and the mean curvature $\alpha=2/R$ are measured from  droplet configurations along the converged string.
Since the free energy $\DF$  is computed by integration, it is known up to a constant and thus can be freely displaced along the $y$-axis. The only free parameter left to compare $\Delta \Omega$ to  $\DF$ is $\sGL$, which should be of the order of the coupling parameter $\Jij$.

Figure~\ref{areasFE}-a shows $\DF$ and $\Delta \Omega$ using $\sGL=1$, which is the same value used for $\Jij$. 
The two quantities show similar trends and are in semi-quantitative agreement for all three substrates analysed in this work (see the \emph{Supporting Information} for the other two substrates).
Interestingly, the ruggedness observed in $\DF$ is also present in $\Delta \Omega$.

In Figure~\ref{areasFE}-b each geometrical term in Equation~\eqref{DOmega} is shown separately and compared with $\DF$. Note the different scales and units for areas (right axis) and $\DF$ (left axis).  
Most of the interfacial areas show a smooth dependence on $f$, the exception being $\ALGbot$ that presents some maximum and minima which is thus a good candidate to explain the roughness of the free energy connected to wetting.

In Figure~\ref{areasFE}-c we show $\ALGbot$ and $\DF$ to assess the importance of the former in the free energy. Vertical dashed lines correspond to maxima of $\DF$. We observe variations in the curve of $\ALGbot$ that are indicated by vertical red lines.  The typical size of these variations is $\Delta \ALGbot \approx 1000L^2$. In terms of dimensionless energy, this corresponds to $\Delta\ALGbot \sGL\approx 1000$.
This energy variation is comparable to the free energy barrier for S$_3$, $\Delta\ALGbot \sGL \approx H^R$, which suggests that the term relative to the interface between liquid and gas below the droplet plays an important role in generating local minima in the free energy.
For S$_1$, instead, abrupt variations of $\ALGbot$ are not sufficient   to generate local minima, due to the steep slope of $\DF$ vs $f$, see the \emph{Supporting Information}.

In the previous section, we mentioned that local minima of $\DF$ correspond to pinning of the drop at the pillars edges, which gives rise to several possible minimal configurations characterised by  different numbers of pillars. To connect this picture with the observation that variations of $\ALGbot$ correlate with  variations of $\DF$, we suppose that the wet domain below the droplet can be approximated by a cylinder of height $h$ and increasing radius $B$, such that $\ALGbot = 2\pi h B$. This approximation does not take into account the roughness of $\ALGbot$ shown in Figure~\ref{ContFE}-c, but is reasonable in the case of minima.
From the volume differences between neighbouring  minima in Figure~\ref{areasFE} we can thus compute the jump in droplet radius $\Delta B$ in the cylindrical approximation, $ \Delta B \approx \Delta \ALGbot /(2\pi h)\approx 16L$. The estimated value of $\Delta B$ corresponds to the  typical size of a cavity $\dd=\ww+\aaa$, which is plausible and supports the idea that local minima correspond to jumps of the droplet front across discrete numbers of pillars.

To summarize, we propose that the global minimum of the free energy corresponds to configurations that minimise the total interfacial energy for a droplet of fixed volume. Local minima, instead, occur in correspondence of abrupt variations of the liquid-vapor interface below the droplet connected with the overcoming of individual pillars.

\subsection{Minima of the free energy and contact angle hysteresis }

We now evaluate the free energy barrier sizes that separate the local minima and the consequences on the metastability of the substrates. 
In Figure~\ref{FE}-d, the barriers $H^R$ and $H^L$ are defined. From Figure~\ref{FE}-a, one measures for S$_1$ typically $H^R \in (3000, 9000)$ and $H^L \in (0, 100)$ showing that, even when the system is initialised in a W-like state, it rapidly evolves towards the CB minimum. For substrate S$_3$ both barriers vary typically in the range $H^R$ $\approx$ $H^L \in (370, 3700)$, which in physical units are between $9 \times 10^{-12}$J and $9 \times 10^{-11}$J, assuming $L=1 ~ \mu$m.
This is much higher than the thermal energy $k_BT \approx 4.1 \times 10^{-21}$~J at ambient temperature: thermal fluctuations are not sufficient to drive the system from one minimum to the other. Only other larger sources of energy can move the drop away from local minima, e.g., mechanical vibrations. 
 
When the system is prepared with some generic initial condition, it will fall in the closest minimum and remain trapped there; this was verified by running  unrestrained MC simulations from several points along the curves in Figure~\ref{FE}-a. Only for the S$_1$ the system always returned to the superhydrophobic CB state, even though two shallow minima were identified numerically; this can either be due to the numerical accuracy of the free-energy profiles or to the size of the barriers, which is so low that the fluctuation imposed by the effective temperature of the Monte Carlo simulations are enough to bypass them.
In other words, the free energy profiles in Figure~\ref{FE} help understanding the origin of contact angle hysteresis in contact angle measurements, which can be rationalised in terms of a rough free energy landscape with multiple minima, separated by large free energy barriers. 
Furthermore, the importance of the initial conditions becomes apparent, which correspond to the preparation of the sessile drop experiment.

\begin{figure}[H]
\centering
\includegraphics[width=.6\linewidth]{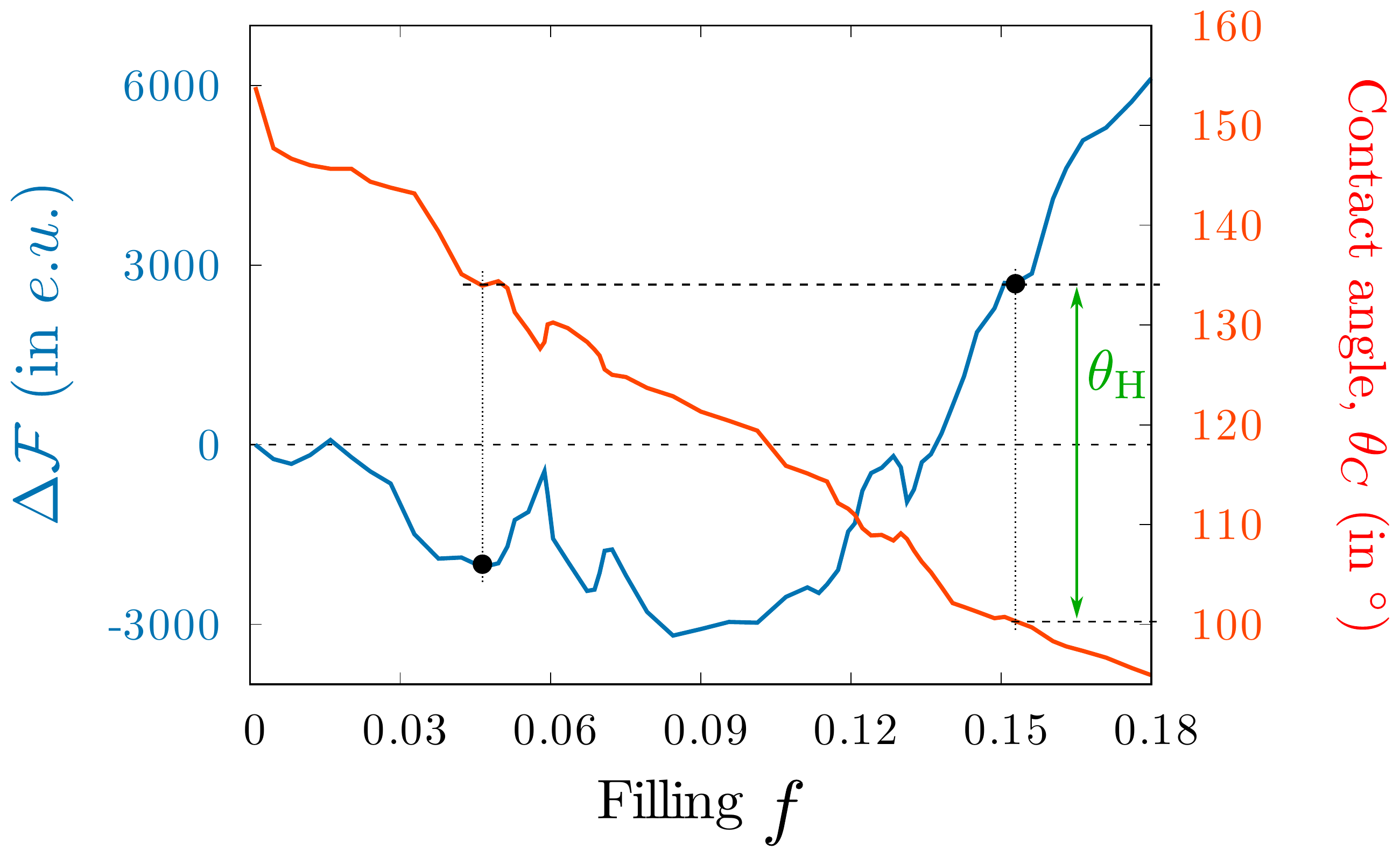}
\caption{Wetting free energy and apparent contact angles as a function of liquid filling for substrate S$_3$. The scale on the left corresponds to $\DF$ and, on the right, to $\ttc$.  The difference between the $\ttc$ associated to the first and last local minimum (indicated as black circles) is defined as the $\tth$.}
\label{FEwCA}
\end{figure}

Finally, the profiles account for the superhydrophobic properties of the CB state, which are connected to the existence of a single minimum, i.e., with low hysteresis. On the other hand, the presence of multiple W minima explains why CAH is so pronounced in this wetting state and its stickiness \cite{Nosonovsky2007,Sbragaglia2007,callies2005}. 
Figure~\ref{FEwCA} shows the apparent contact angle $\ttc$ of the droplet measured in MC simulations together with $\DF$ for substrate S$_3$.
The fact that there is a basin of Wenzel states and that each minimum has a different contact angle  allows us to define the contact angle hysteresis $\tth$ related to the wet W state  as the difference in $\ttc$ of the configuration associated to the first and the last local minimum of $\DF$, see Figure~\ref{FEwCA}.  Table~\ref{tab1} summarizes the values $\tth$ for the three substrates together with their roughness ratio $r$. We measured an increase of $\tth$ when $r$ decreases, which is in line to what was previously observed \cite{wang2020}. 

\begin{table}[H]
\centering
\begin{tabular}{| c || c | c||} 
\hline
Substrate &  roughness ratio, $r$ & $\tth$, (in $^{\circ}$) \\ [0.5ex] 
\hline\hline
S$_1$ & 3 & 0  \\ 
\hline
S$_2$ & 2.2 & $24^{\circ}$ \\
\hline
S$_3$ & 1.8 & $36^{\circ}$ \\ [1ex] 
\hline
\end{tabular}
\caption{Contact angle hysteresis $\tth$ and roughness ratio  $r$ for the considered substrates.}
\label{tab1}
\end{table}

\section{Discussion: modeling rough wetting}
\label{sec:discussion}
We have identified by free energy simulations that some pillared substrates present several local minima separated by high barriers, while others  present only one minimum. Incidentally, for tall/tightly packed pillars  only the superhydrophobic CB state is possible, while for more sparse pillars multiple local wet minima arise. 
The latter surfaces thus display a behavior which is strongly dependent on the initial conditions: if a droplet is deposited on a substrate at a random configuration, it would accommodate in a wetting state correspondent to the closest minimum which can be either superhydrophobic or sticky. The present findings also imply that theories that predict a single W state cannot be complete \cite{Quere2008}. 

\begin{figure}[H]
\centering
\includegraphics[width=.6\linewidth]{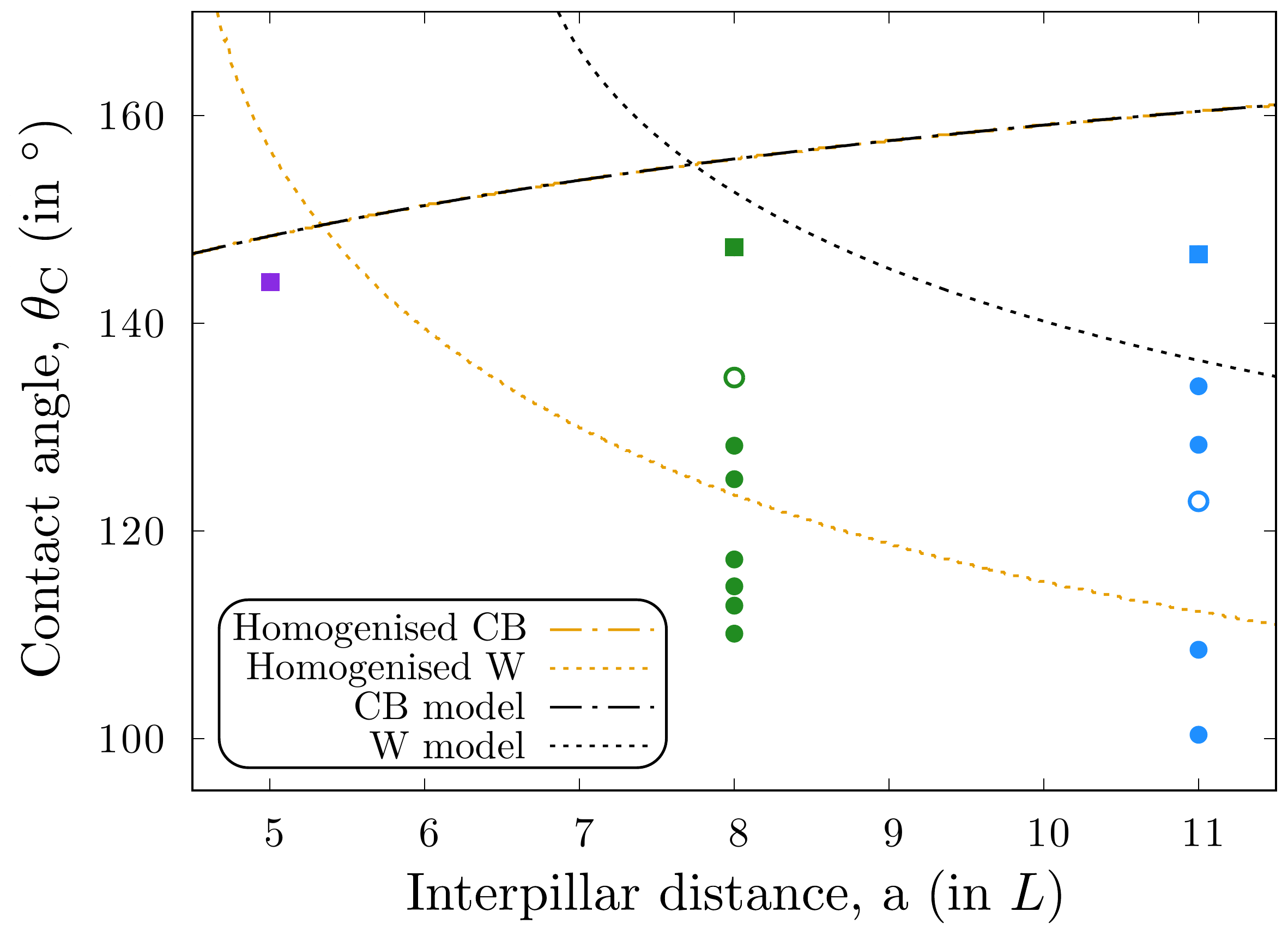}
\caption{Apparent contact angle computed at physical minima in Figure~\ref{FE} as a function of the distance between pillars. Squares are $\ttc$ of the CB state, circles for the W minima; the open circles denote the global W minimum. Orange lines are the solution for the homogenised models: dot-dashed for CB, Equation~\eqref{en_CB}, and dotted for W, Equation~\eqref{en_W}. Black lines are the predictions of the classical models: dot-dashed for CB, Equation~\eqref{angleCB}, and dotted for W, Equation~\eqref{angleW}.}
\label{CA_simuModels}
\end{figure}

We measured all apparent contact angles $\ttc$ of the droplet in configurations correspondent to physical minima identified in our simulations for the three substrates and compared them with theoretical models, see Figure~\ref{CA_simuModels}. Squares corresponds to the superhydrophobic CB  minimum and circles to wet  W states. Lines are solutions of the classical Cassie-Baxter and  Wenzel models, whose apparent contact angles are  given by:
\begin{eqnarray}
\cos\ttc^{\textrm{CB}} &=& \phiS \cos\tty - (1-\phiS),\label{angleCB}\\
\cos\ttc^{\textrm{W}} &=&  r \cos\tty \;.\label{angleW}
\end{eqnarray}

For the case where the global minimum is CB, which happens for substrate S$_1$, Figure~\ref{CA_simuModels} shows that the prediction of the  Cassie-Baxter model is almost quantitative. The CB model also has a reasonable agreement with simulated $\ttc$ in cases where it is only a local minimum, which happens for substrates S$_2$ and S$_3$. The agreement with CB model deteriorates as the pillar distance increases, which can be explained by the increasing curvature of the menisci suspended among pillars, not accounted for in the classical CB model.

In principle, no direct comparison can be made between the W model and the MC results, mainly because the model predicts only one state, while simulations demonstrate the existence of multiple wet states. However, a fair comparison can be made considering only the global minimum, which is compatible with the minimisation procedure used in all homogenised models. It is seen that the W model prediction is far from the measured contact angle and, for S$_2$, it even predicts a contact angle higher than the actual CB one.
When the finite  size of the droplet is taken into account, as done in Equation~\eqref{en_CB} and \eqref{en_W}, the solution for the CB does not change but the W curve shifts to smaller values. This simple correction captures the overall trend with interpillar distance and improves the agreement with the global minima, although it is not quantitative.  However, we remark that the finite size is not enough to account for the existence of multiple minima, which is due to pinning at individual pillars and is crucial to account for contact angle hysteresis.  

Figure~\ref{CA_simuModels} confirms that the extent of CAH has the following trend  $\mathrm S_1<\mathrm S_2<\mathrm S_3$, which suggests that the more favorable the W state, the higher $\tth$. This trend is related to the number of local minima and to the facility of wetting the bottom of the surface for short pillars, but what limits the total number of minima still remains an open question.

\section{Conclusions} 
\label{sec:conclusions}

In this work, we have used a combination of the cellular Potts model~\cite{Graner1992, Fernandes2015} and the string method~\cite{e2002,maragliano2006,e2007} to compute numerically the free energy $\DF$ of a droplet with a fixed volume placed on  surfaces decorated with pillars. The computed free energy landscape for wetting the surface is found to be rough, with one minimum corresponding to the superhydrophobic Cassie-Baxter state; depending on the pillar spacing, multiple local minima can exist, associated with different wet Wenzel states. Free energy barriers typically larger than the thermal energy are found, whose origin can be traced back to pinning of the liquid-gas interface below the droplet at individual pillars. Wenzel minima are characterized by an increasing number of filled cavities, corresponding to different apparent contact angles. This scenario accounts for the strong contact angle hysteresis of the W state(s), in which the final wetting state depends on the initial condition. 
 
We have compared the apparent contact angles obtained in our \emph{in silico} experiments with the predictions of simple models, including the classical Cassie-Baxter~\cite{Cassie1944} and Wenzel~\cite{Wenzel1936} ones. Results showed that the Cassie-Baxter model has a good prediction capacity, which can be improved by considering the curvature of menisci overhanging surface textures. On the other hand, the prediction of the apparent contact angle cannot be made by simple models in the W state effectively hindering the deduction of surface characteristics from $\ttc$ measurements. Actually, the problem of predicting $\ttc$ in the W state seems ill-posed and the relevant challenge is to assess its contact angle hysteresis, which can be achieved only by detailed models.

An interesting and long-standing question is the possibility of predicting CAH~\cite{joanny1984, reyssat2009, giacomello2016wetting}, which was possible for individual surfaces within our approach.  It is found that the wider the pillars are spaced, the larger CAH. Our analysis suggests that larger drops would have liquid enough to fill more cavities and thus the free-energy profile would show even more local minima. What limits the total number of filled cavities and how these local minima connect to CAH is however not clear and deserves to be investigated.

Finally, our findings suggest that, for a complete characterization of the wetting properties of a substrate, it is necessary to repeat sessile drop experiments several times for different initial conditions. Moreover, the fact that free energy barrier between minima are typically much larger than thermal fluctuations suggests that mechanical vibrations of the substrate are necessary to drive the droplet across different minima. The amount of mechanical energy necessary to reach different minima need to be evaluated by computing the barrier sizes, which is possible with the methodology proposed in this work.



\medskip
\textbf{Supporting Information} \par 
The supporting information describes the algorithm used to find the thermodynamic wetting state for droplet placed on the pillared surface. Its also presents the criteria to adjust parameter $\kappa$ and description of the videos available. Then there is a discussion of the comparison between the free energy obtained from the simulation with the Cassie-Baxter and Wenzel models. Finally the areas of the components for the proposed analytical free energy are presented.

\medskip
\textbf{Acknowledgements} \par 
M. Silvestrini and C. Brito thanks CAPES and CNPq for financing this work.  
This work has been supported by the project ``Understanding and Tuning FRiction through nanOstructure Manipulation (UTFROM)'' funded by MIUR  Progetti di Ricerca di Rilevante Interesse Nazionale (PRIN) Bando 2017 - grant 20178PZCB5.
This research is part of a project that has received funding from the European Research Council (ERC) under the European Union’s Horizon 2020 research and innovation programme (grant agreement No. 803213).

\medskip

%
\bibliographystyle{MSP}
\bibliography{droplet_int}

\end{document}